# On the Value of Project Productivity for Early Effort Estimation

**Mohammad Azzeh[1]**
Department of Data Science
Princess Sumaya University for Technology, Amman, Jordan
m.azzeh@psut.edu.jo

**Ali Bou Nassif**
Department of Computer Engineering
University of Sharjah
Sharjah, UAE
anassif@sharjah.ac.ae

**Yousef Elsheikh**
Department of Computer Science
Applied Science Private University, Jordan
y_elsheikh@asu.edu.jo

**Lefteris Angelis**
School of Informatics
Aristotle University of Thessaloniki
Thessaloniki, Greece
lef@csd.auth.gr

**Abstract.** In general, estimating software effort using a Use Case Point (UCP) size requires the use of productivity as a second prediction factor. However, there are three drawbacks to this approach: (1) there is no clear procedure for predicting productivity in the early stages, (2) the use of fixed or limited productivity ratios does not allow research to reflect the realities of the software industry, and (3) productivity from historical data is often challenging. The new UCP datasets now available allow us to perform further empirical investigations of the productivity variable in order to estimate the UCP effort. Accordingly, four different prediction models based on productivity were used. The results showed that learning productivity from historical data is more efficient than using classical approaches that rely on default or limited productivity values. In addition, predicting productivity from historical environmental factors is not often accurate. From here we conclude that productivity is an effective factor for estimating the software effort based on the UCP in the presence and absence of previous historical data. Moreover, productivity measurement should be flexible and adjustable when historical data is available.

## 1. Introduction

Estimating software effort in the early stages of the software development cycle is a critical task of software project management[1]–[4]. Managers usually need such early estimates in order to bid on a project contract and make appropriate planning decisions [5]–[8]. Since the available information is often not sufficient, the estimation process produces a result characterized by a large degree of uncertainty[1]. To mitigate this issue, managers use functional requirements as input to provide early effort estimates. Software project size provides a general sense of how large a software project and thus gives an indication of how many resources the project will need to achieve successful results. Functional size measurement, such as Function Points (FP) or Use Case Points (UCP), is a major driver for early effort estimation methods. Although the construction of FP has been widely discussed and questioned, it remains the most popular sizing technique in cost estimation methods [9]–[11]. However, counting FP requires special care, particularly with regard to assessments of complexity and environmental factors [12]. On the other hand, Karner [13] proposed an FP-like method called UCP that uses a use case diagram as an input to estimate the potential size of object-oriented software projects at the requirements phase. The philosophy behind UCP is inspired by both FP and MarK II FP. The UCP method is relatively easy to use, but an experienced software estimator is required to translate the use case diagram elements into their corresponding size metrics [12]. The total UCP is obtained through the steps of the calculations and its accuracy depends on the degree of detail of the use case diagram. Although the use case diagram is a de-facto UML analysis diagram, the method has not been widely adopted among software organizations due to the lack of consensus on how to use UCP to determine effort [1].

The productivity variable is often used as a key factor in producing effort estimates with UCP when the organization does not have a historical dataset [13], [14]. Software productivity is defined as the ratio between the quantity of software produced (i.e.: size), labor and production cost (i.e.: effort) [15]. However, this definition is often misleading because the quotient between size and effort is interpreted as productivity, while the opposite quotient is known among researchers as Project Delivery Rate. The second interpretation is the most commonly used type of data in UCP effort estimation [13], [16], [17]. For simplicity, the term productivity is used throughout this paper, except for Nassif model where productivity is interpreted as size/effort [6].

---

[1] Corresponding Author.

The outstanding question is how to find an optimal productivity value that accurately determines the UCP size metric to effort. Karner suggests using the default productivity value, which is equal to 20 person-hours per UCP [13]. Karner found that all the projects examined tended to use the same value of productivity. This assumption became the basis for recent models, specifically when no historical data set was available. Several studies subsequently published that criticized this assumption and showed that using the same value of productivity for all projects - regardless of type, complexity, and team size - is meaningless and harms accuracy of the predictions[6], [17]–[20]. In light of this, Schneider and Winters (S&W) [17]recommended the use of three levels of productivity that are assigned based on environmental factors. They stated that these environmental factors must be evaluated first because they affect the efficiency of the project team. This approach uses three predetermined levels of productivity, but these levels are not flexible and cannot be adjusted for the project on a case-by-case basis. Likewise, Nassif et al. [6]proposed four levels of productivity ratios to be used in their model. The productivity is adjusted by computing the sum of the products of environmental factors and their degrees of influence and converting them into productivity ratios using a fuzzy model.

However, in practice, productivity is difficult to be measured because it is affected by many factors including the amount of code reused, the type of model used in the project, the effectiveness of communication for the development team and project deliverables. While productivity may increase when using development best practices, this is not guaranteed, due to circumstances beyond the development team's control. So, productivity should be measured based on the value of results to the software consumer, and not just based on the size of a software product [21]. From the previous studies, we can determine the following points: (1) Researchers consider productivity to be the second variable in computing effort [6], [15], [16], [22]. (2) Studies are generally based on fixed productivity ratios or a very limited range of ratios, which is not an effective approach [18]–[20]. (3) The current research was not succeeded in learning and adjusting productivity based on a historical dataset [23]. However, we believe that when we have the data to do so, we should be able to compute productivity using flexible and adjustable methods. Flexibility means that the productivity prediction must be affected by UCP factors, and adjustability means that the productivity of any given project must be adjusted based on the productivity of similar historical projects. The large software repository made available recently allows for further empirical research. Since little work has been done to validate the above-mentioned issues, we are motivated to empirically investigate the impact of productivity on UCP effort estimation. Thus, the present paper will address the following research questions:

**RQ1**: Can learning productivity values from historical datasets improve the accuracy of UCP effort estimation?
To address RQ1, we first examined the distribution of the productivity variable in each dataset used. Multiple experiments were conducted to empirically compare between effort estimation models using UCP and those using productivity as data inputs. In addition, we examined another method that predicts and adjusts productivity: analogy-based estimation with Regression to Mean (hereafter called R2M) [16], [24]. This model was previously used in software effort estimation in the work of Jorgensen et al. [16] to adjust productivity based on FP. The R2M method allows us to learn and adjust productivity dynamically, using data from similar previous projects.

**RQ2**: Is there a strong relationship between productivity and summation of environmental factors?
To address RQ2, we analyzed the relationship between the productivity variable and other variables in the datasets. The aim of this analysis is to find if there is way that can help practitioners to estimate productivity from environmental factors rather than guessing or relying on not related factors.

**RQ3**: Does using a homogeneous data set improve the accuracy of estimating productivity?

To address RQ3, we divided the datasets into smaller and more homogenous datasets and then applied each of the four models to these datasets. It is well recognized the working on homogenous data (i.e., local data) can improve prediction accuracy of effort estimation [3]. Therefore, we need to investigate if this issue has positive impact on the performance of the productivity prediction or not.

The remainder of the paper is structured as follows: Section two introduces the UCP method. Section three presents related work. Section four discusses the preparation and methodology of the experiment. Section five describes the dataset and provides an analysis of the productivity measures. Section six presents the results we obtained from our analysis. Section seven presents the discussion of the results. Section eight introduces the threats to validity of our experiment and finally the paper ends with the conclusion.

## 2. Overview to Use Case Points

The UCP method is used to estimate the size of object-oriented software projects at early phases of software development [13]. The original UCP model consists of four main metrics: (1) Unadjusted Actor Weights (UAW), (2) Unadjusted Use Cases (UUC), (3) Technical Complexity Factors (TCF) and (4) Environmental Factors (EF). The UAW is measured by counting and classifying the actors into three levels (simple, average, and complex) according to complexity level [25], [26]. Each of these categories is multiplied by a predefined weight as shown in Equation 1.

$$UAW = 1 \times \#Simple\_Actors + 2 \times \#Average\_Actors + 3 \times \#Complex\_Actors \quad (1)$$

The UUC is calculated by counting and classifying use cases into three categories (simple, average and complex) according to how many transactions occur in the use case, as shown in Equation 2. A simple use case is one in which there are three or fewer transactions; an average use case has between four and seven transactions; and a complex use case has more than seven transactions. In this case, transactions are seen as a stimulus and response occurring between an actor and the system. Each use case type is associated with a predefined weight.

$$UUC = 5 \times \#Simple\_UseCases + 10 \times \#Average\_UseCases + 15 \times \#Complex\_UseCases \quad (2)$$

The TCF is calculated from a set of 13 factors as shown in Equation 3, where each factor has a value between 0 and 5 and is adjusted according to its associated weight.

$$TCF = 0.6 + \left(0.01 \times \sum_{i=1}^{13}(f_i \times fw_i)\right) \quad (3)$$

where $f_i$ is the value of influence of factor $i$, and $fw_i$ is the weight associated with factor $i$.

The Environmental Factors (EF), as shown in Equation 4, are calculated from a set of eight factors where each factor has a value between 0 and 5 and is adjusted according to its associated weight.

$$EF = 1.4 - \left(0.03 \times \sum_{i=1}^{8}(env_i \times ew_i)\right) \quad (4)$$

where $env_i$ is the value of influence of factor $i$ and $ew_i$ is the weight associated with factor $i$.

Finally, the UCP value is calculated as shown in Equation 5.

$$UCP = (UAW + UUC) \times TCF \times EF \quad (5)$$

The UCP value is then translated to human effort in man-hours by multiplying it with the predicted productivity as shown in equation 6.

$$Effort = Productivity \times UCP \quad (6)$$

## 3. Related Work
### 3.1. Use Case Points

The original Use Case Points (UCP) model was first proposed by Karner in 1993 [13]. Since then, some researchers have attempted to improve the UCP model [5], [12], [14], [23], [27]–[34]. In an attempt to validate the UCP model, others have compared it with the FP model [25], [35]–[41]. Many studies have aimed to develop predictive models based on UCP[20], [42]–[44].

In our previous Systematic review study [23], we described the main research directions in UCP method which are divided into three main directions: (1) building effort prediction models from UCP method, (2) improving the structure of UCP method and (3) comparisons with other functional size measures. In this paper we focus on the studies that build effort prediction models based on UCP. Other studies that focus on improving structure of UCP method are out of scope in this study.

In the related literature, the most advanced effort estimation models use fuzzy systems based on UPC components [26], [44]. For example, Nassif et al. [6] used Mamdani and Sugeno fuzzy models to calibrate the abrupt changes in use case complexity weights. The authors also proposed a neuro-fuzzy model to calibrate the weights. Pantoni et al. [45]used fuzzy theory to estimate the effort for original software tasks based on the UCP method. The fuzzy system used for this approach requires three inputs; namely, the complexity of the modules, the developers' experience, and the size of the alteration. It uses this data to estimate the effort time needed to perform a given task. The proposed model may prove to be a good alternative to other estimation methods used when no historical data is available. When estimating long-term effort, however, other techniques produce more accurate results. Iraji and Motameni [44] proposed an adaptive fuzzy neural network model to predict the effort of object-oriented software projects based on UCP. Kamal et al. [46], [47]used a fuzzy logic system in an estimation process that takes into consideration the imprecise information that is available in the early stages of software development. The authors also proposed a classification framework that compares various approaches based on use case points, thereby providing valuable information to practitioners.

Neural network-based models have also been used to estimate software effort in a given system [48]–[50]. These neural network models require data inputs such as software size in UCP and other qualitative attributes such as complexity. A Treeboost model was used to predict software effort based on UCP software size and software team productivity [51].

Various studies have also used regression models, with non-linear regression models being the most frequently used [6], [33], [34], [52], [53]. Frohnhoff and Engels [54] applied a UCP model on 15 industrial projects and found that the results deviated considerably. They suggested improving the UCP method by standardizing the weighting of use cases and providing a user guide that defines the level of abstraction and complexity of a use case.

## 3.2. Software Productivity

Research on the measurement of software productivity is quite limited, with most studies focusing on comparing effort and other size measures. Kitchenham and Mendes [15] observed that productivity could be measured using a variety of size measures. They hypothesized that size and effort were related and conducted an experiment in which reused components were not included. Jorgenson et al. [16] applied regression towards the mean to predict productivity based on data from previous projects in case-based reasoning. This approach is especially useful for datasets that have been divided into a subset of homogenous projects. Rodriguez et al. [22] examined team size and productivity variables in the International Software Benchmarking Standards Group (ISBSG) dataset with an objective to examine their impact on effort prediction. They measured these variables after dividing the dataset into smaller categories based on development type and programming language used. They concluded that the productivity of a team composed of nine or more members decreases, and that projects aiming to enhance existing software record better productivity than new projects. Sentas et al. [55]used Multinomial Logistic Regression to estimate productivity based on independent software variables. Petersen [56] conducted a systematic review of various methods of software productivity measurement. Premraj et al. [57] investigated how software productivity has changed over time. They found that productivity has improved over time and can vary based on business domain. Azzeh et al. [5], [18], [20]investigated the use of productivity along with environment factors to improve early phase predictions. They found environmental factors to be good predictors for productivity when historical data is absent.

The use of productivity as an input in UCP effort estimation has not been thoroughly investigated in the literature. Most productivity values recognized in academic studies were suggested by experts or as a result of small experimentation. Karner suggested using 20 person-hours/UCP. Schneider and Winters [17] proposed three levels of productivity (fair, low, very low) that were designated through an analysis of environmental factors. They assigned 20 person-hours/UCP for the fair group, 28 person-hours/UCP for the low group and 36 person-hours/UCP for the very low group. Alves et al. [58]found a discrepancy in the number of hours required to develop one UCP between the original model (20 hours / UCP) and their proposed model (between 6 and 13 hours / UCP). Others found several issues in the design of UCP, looking into the explicit and implicit principles of UCP, the correspondence between measurement and empirical reality and the consistency of its system of points and weights [18], [20]. Nassif et al. [6]proposed four productivity ratio levels. In the study, productivity was computed with a fuzzy system based on a summation of environmental factors. A log-linear regression model was then applied to the productivity and UCP data to calculate software effort.

In summary, the conclusions drawn from these studies are: (1) The use of fixed productivity ratios, or a limited range of ratios, in the original model and its subsequent variants did not produce effective results for the software industry. (2) When historical data is available, very few studies investigate the validity of using productivity as input data in UCP effort prediction. (3) There is a lack of research on learning and adjusting productivity based on historical data. In this paper, we attempt to address these challenges by using a large dataset sourced from the industry and education sectors. We also address the problem of productivity by conducting an empirical comparison of four different effort models.

## 4 Experimental Setup
### 4.1 Choice of productivity prediction methods

The purpose of this study is to investigate the effect of productivity in predicting effort early in project development when using the UCP sizing technique. To achieve this, we used three datasets and selected four software effort estimation models from the literature that used productivity as a cost driver as well as a size variable. The models selected are the Karner model, the Schneider and Winters model (S&W) [17], the Nassif model [6] and the Regression to Mean model (R2M) [16]. The Karner model uses constant productivity values for all projects, regardless of type and complexity. The Schneider and Winters model (S&W) uses three levels of productivity (20, 28 and 36 person-hours) that are allocated according to the analysis of environmental factors as described in Equation 7. The main idea is to count the number of factors from 1 to 6 with influence values less than three, as well as the number of factors from 7 to 8 with influence values larger than three. The efficiency is then computed on the basis that data. If the total efficiency value is less than or equal to 2, productivity is considered to be 20 person-hours. If the total count is between 3 and 4 (inclusive), productivity is considered to be 28 person-hours. Finally, if the total count is greater than 4, then the productivity is considered 36 person-hours.

$$Effort = \begin{cases} 20 \times UCP & total\_Count \leq 2 \\ 28 \times UCP & 3 \leq total\_Count \leq 4 \\ 36 \times UCP & total\_Count > 4 \end{cases} \qquad (7)$$

The third approach is the Nassif model [6], which constructs a non-linear relationship between UCP, effort, and productivity as shown in Equation 8. In this case, productivity is the quotient between size and effort. It assumes that the relationship between effort and UCP follows a non-linear pattern; in fact, it is supposed to follow an exponential curve. The factors α and β are determined by the analysis of the training dataset. In the original model, the values were α=8.16 and β=1.17. For purposes of replication, we are unable to use the same values due to changes in the training datasets as a result of leave-one-out cross-validation. Therefore, at each iteration, we compute the values for each training dataset separately. The productivity factor is computed based on $prod\_sum = \sum_{i=1}^{8}(env_i \times w_i)$ from the data based on environmental factors. Then we use this in their proposed fuzzy model which converts the $prod\_sum$ in to a value representing productivity based on the following rules:

1- If $prod\_sum$ is less than 0, then productivity = 0.4.
2- If $prod\_sum$ is between 0 and 10, then productivity = 0.7.
3- If $prod\_sum$ is between 10 and 20, then productivity = 1.
4- If $prod\_sum$ is greater than 20, then productivity = 1.3.

$$Effort = \frac{\alpha}{Productivity} \times UCP^\beta \qquad (8)$$

In addition to the above-mentioned models, we use the Regression to Mean (R2M) model to predict productivity. Jørgensen et al. [16] was the first researcher to apply R2M to the area of software effort estimation by integrating it with analogy-based estimation. The method assumes that if similar projects have extreme productivity values, then the productivity value of the project under analysis must be adjusted to approximate the average productivity value of the projects in the training data set. The R2M model is used to predict the effort of new projects based on the productivity of training projects, as shown in Equation 9. The nearest neighbor is found using Euclidean distance and based on the eight environmental factors ($env_1$ to $env_8$).

$$Effort = \left(PDR_c + (h - PDR_c) \times (1 - r)\right) \times UCP \tag{9}$$

where $PDR_c$ is the productivity of the nearest analogy, $h$ is the average productivity of the training projects, and $r$ is the historical correlation between non-adjusted analogy-based productivity and actual productivity.

The last approach is to use Naïve method such mean or median of the productivism of the training dataset. We first investigate the distribution of productivity in the training dataset. If it is normal we use mean, otherwise we use median. This approach is called Naïve.

## 4.2 Empirical Validation Strategy

It is clear from the above descriptions that the Karner model and the S&W model can be used even without historical project data. So, there is no need to split the datasets into training and testing. The UCP prediction model is straightforward, while the S&W model depends on the analysis of environmental factors. The remaining models (Nassif, R2M and Naïve) require historical project data and cannot be validated without training and testing data. To validate these two models, the leave-one-out cross-validation method was used. In each instance, one project is selected as a test set, while the others are used as a training set. The model uses training data to construct the prediction model, while testing data is used to validate the model. This procedure is performed until all projects within the testing dataset are used as test projects. During each iteration of the model, the accuracy of each test run is checked, and the evaluation is recorded.

## 4.3 Evaluation measures

Indeed, a substantive discussion has taken place regarding the suitability of evaluation measures and their biases. The most commonly used evaluation measures, such as the Magnitude of Relative Error (MRE) and its derived measures, the Mean Magnitude Relative Error (MMRE) and the Performance indicator (*pred*) have been criticized for being biased and unbalanced in many validation circumstances due to the asymmetrical distribution produced by the MRE[59], [60]. To avoid this drawback, we used the Mean Absolute Error (MAE) evaluation measure, which is unbiased and does not present asymmetrical distribution[59]. The MAE is calculated by taking the average of Absolute Error (AE) as shown in Equations 10 and 11. In addition to MAE, we used two other measures of accuracy reported in the literature [59] which are less prone to bias or asymmetric distribution. The two measures are Mean Balanced Relative Error (MBRE) and Mean Inverted Balanced Relative Error (MIBRE), as shown in Equations 14 and 15, respectively.

To ensure that all models are predictive and not guessing, we use two validation measures proposed by Shepperd and MacDonell [59]. The first accuracy measure is called Standardized Accuracy (SA), as shown in Equation 12. The second measure is the effect size shown in Equation 13. However, the SA measure is mainly used to test whether or not the prediction model under evaluation really outperforms the baseline of random guesses, thus generating meaningful predictions. If this baseline accuracy value is not surpassed, we cannot claim that the prediction model is effective. SA can be interpreted as a measure of improving of random guessing for a given model; it provides an idea of how well the approach works. Effect size ($\Delta$) is used to ascertain whether or not the predictions of the model under evaluation are generated by chance, as well as to determine how well the model performs better than random guessing. This test should be done because the statistical significance test is not informative when the two prediction models are different. The value of $\Delta$ can be interpreted as a function of the categories small (0.2), medium (0.5) and large (0.8), where values greater than or equal to 0.5 are considered more efficient. If the effect size is greater than 1, the difference between the two means is greater than one standard deviation. A result greater than 2 means that the difference is greater than two standard deviations.

$$AE_i = |y_i - \hat{y}_i| \tag{10}$$

$$MAE = \frac{\sum_i^n AE_i}{n} \tag{11}$$

$$SA = 1 - \frac{MAE}{\overline{MAE}_{po}} \tag{12}$$

$$\Delta = \frac{MAE - \overline{MAE}_{po}}{SD_{po}} \tag{13}$$

$$MBRE = \frac{1}{n}\sum_{1}^{n}\frac{AE_i}{min\ (y_i, \hat{y}_i)} \tag{14}$$

$$MIBRE = \frac{1}{n}\sum_{1}^{n}\frac{AE_i}{max\ (y_i, \hat{y}_i)} \tag{15}$$

Where:
 $y_i$ and $\hat{y}_i$ are the actual and estimated effort of a project.
$MAE$ is the mean absolute error of the estimation model.
$\overline{MAE}_{po}$ is the mean value of a large number of runs of random guessing. This is defined as: predict a $\hat{y}_i$ for target case $t$ by randomly sampling (with equal probability) over all the remaining $n$ - 1 cases and take $y_t = y_k$ where $k$ is drawn randomly from $1…n \wedge k \neq t$. This randomization procedure is robust, as it makes no assumptions and requires no knowledge concerning population.
$SD_{po}$ is the sample standard deviation of the random guessing strategy.

## 5  Dataset and Productivity Analysis
### 5.1 Dataset description

Previous models in UCP were tested on a very limited number of projects, which in turn reduced the credibility of the results. To mitigate this, the datasets used in this study come from two main sources: Ochodek et al. [12] and Nassif et al. [6]. The first source contains a single dataset of 14 projects collected from industry and student projects. Half of these projects were developed in a company, while five projects were developed at the Poznan University of Technology as student projects. The remaining two projects were developed as "student to business" projects. Some projects were developed from scratch, while others were customized for new customers. This dataset was mainly used in the study of Ochodek et al. [8] to investigate the effect of UCP components (i.e., UAW, UUC, EF and TCF) on effort estimation. Table 1 shows the statistical characteristics of these projects.

The other source includes two homogenous datasets. The first dataset (hereafter referred to as "Industrial") was collected from completed industry projects, while the second dataset (hereafter referred to as "Education") was collected from university student projects. The Industrial dataset contains 45 projects developed from information systems projects with clients including hotel chains, multi-branch universities and multi-warehouse bookstores. The architectures used in developing these projects consist of 2-tier desktop applications and 3-tier web architectures. The Education dataset contains 65 projects collected from 4[th] year software engineering students at a university in Canada. These senior design projects are developed and implemented using UML and object-oriented languages. Cuauhtémoc et al. [61] stated that studies should not ignore small projects. Predicting the effort of developing student programs is helpful in many disciplines such as education. Moreover, small programs developed by one developer (such as projects in an Education database) are of interest to many practitioners, as confirmed by Personal Software Process (PSP) [62]. Tables 2 and 3 present the descriptive statistics for the two datasets.

Table 1: Descriptive statistics of the Ochodek dataset

| Variable | Mean | StDev | Min | Median | Max | Skewness | Kurtosis |
|---|---|---|---|---|---|---|---|
| UCP | 88.3 | 70 | 22 | 75 | 304 | 2.51 | 7.45 |
| Effort | 1266 | 1002 | 277 | 958 | 3593 | 1.37 | 1.24 |
| Productivity | 15.07 | 7.5 | 4 | 14 | 35 | 1.32 | 3.15 |

Table 2:  Descriptive statistics of the Industrial dataset

| Variable | Mean | StDev | Min | Median | Max | Skewness | Kurtosis |
|---|---|---|---|---|---|---|---|
| UCP | 739 | 1564 | 33 | 154 | 7027 | 3.14 | 9.85 |
| Effort | 20573 | 4732 | 570 | 3248 | 224890 | 3.26 | 10.69 |
| Productivity | 24.089 | 5.116 | 14 | 24 | 33 | -0.03 | -0.74 |

Table 3: Descriptive statistics of the Education dataset

| Variable | Mean | StDev | Min | Median | Max | Skewness | Kurtosis |
|---|---|---|---|---|---|---|---|
| UCP | 82.6 | 20.71 | 40 | 81 | 149 | 0.85 | 1.26 |
| Effort | 1672.4 | 414.3 | 696 | 1653 | 2444 | -0.05 | -0.82 |
| Productivity | 20.8 | 4.777 | 11 | 21 | 32 | 0.19 | -0.26 |

Figures 1 to 3 summarize three main variables using boxplots. Figure 1 shows the distribution of UCP variable across three benchmark datasets. We can notice that there is a large variability in UCP for Industrial dataset which confirms that the used projects vary in size. The UCP distribution for other datasets is quite compact and less scattered. Furthermore, the UCP distribution of Education dataset is relatively symmetric. Whereas the UCP distribution for Industrial and Ochodek datasets is skewed to right and left respectively. The same interpretation can be also found for the distribution of Effort variable as shown in Figure 2. The different things that we notice here are the variability of Effort variable for three dataset is high, and the distribution of Effort for Industrial and Ochodek is skewed to right. For the distribution of productivity variable in Figure 3 we can notice that the variability is quite similar, but with significant skewness. Further details about productivity variable is deeply explained in section 5.2.

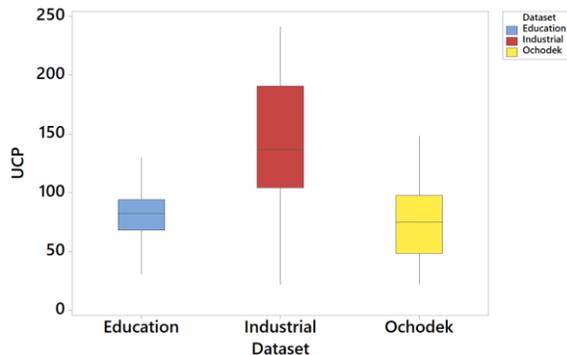
Fig. 1. Boxplot of UCP distribution across three benchmark datasets

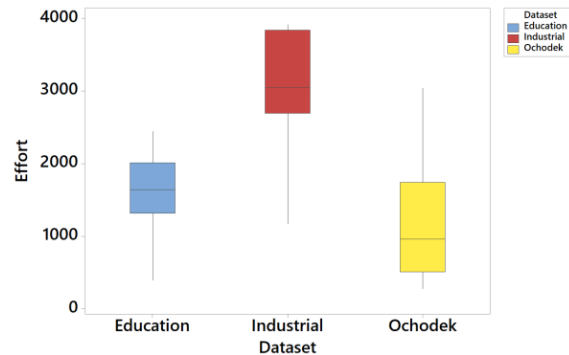
Fig. 2. Boxplot of Effort distribution across three benchmark datasets

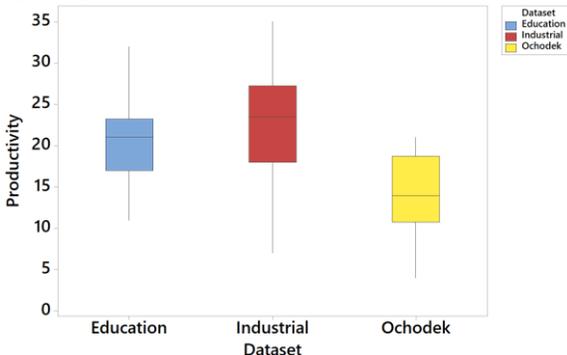
Fig. 3. Boxplot of Productivity distribution across three benchmark datasets

## 5.2 Productivity Analysis

This section presents our analysis of the productivity variable in each dataset. First, we drew the histogram of actual productivity for all datasets, as shown in Figures 4 to 6. The first observation is that although the range of productivity is quite large, especially for the Ochodek dataset, outliers were not detected using the Grubb's test method. As seen in the histograms, the productivity variables are quite normally distributed, although they are skewed either to the right or to the left, which is also confirmed by the skewness measure in Tables 1 to 3. Skewness values between -1 and 1 represent a symmetric distribution. Projects skewed to the right have lower productivity than those skewed to the left. If we look closely at the Ochodek dataset, we notice that productivity is skewed to the right and its distribution is more sharply peaked than normal. The kurtosis is 3.15, as shown in Figure 4 and Table 1. This is an indication that these projects require low productivity. The productivity distributions for both the Industrial and the Education dataset are flatter than normal and have kurtosis values of -0.74 and -0.26. Interestingly, the productivity in the Industrial dataset follows a normal distribution with skewness very close to zero, while the curve is slightly skewed to the right in the Education dataset, as shown in Figures 5 and 6.

Although the data are normally distributed in the histograms, this was also confirmed through the Kolmogorov–Smirnov test for normality when we evaluated the datasets with parametric tests. The peak of the productivity variable varies from one dataset to another, indicating that using the default productivity value in generating effort estimates is not reasonable and most likely leads to inaccurate estimates. Upon further investigation, some datasets have two peaks, indicating that the evaluation would be more effective if the dataset was divided into two groups. One important aspect to note is the variability of productivity spread. The spread of the Ochodek dataset is larger than the Education and the Industrial datasets, which have relatively similar spread to each other. In our comparison between the Education and Industrial datasets, we noticed that the average productivity of the Education projects is lower than that of the Industrial projects. However, the distribution of productivity for the Education dataset is smaller than that of the Industrial projects, probably due to the type of projects developed in the university as graduation projects (which are usually constrained by short timeframes and have lower complexity and lower quality).

Next, we study the relationship between productivity, UCP and effort in each dataset as shown in Table 4. We also measure the correlation between productivity and environmental factors (EF), because this metric is considered as an indicator for productivity in some previous studies. Our analysis was based on Pearson's correlation method. Although there is some degree of association between productivity and effort, it is not a strong correlation. In a similar way, there is no strong correlation between productivity and UCP. However, we observed that EF is closely correlated with productivity on two datasets, confirming the assumption made in some studies that environmental factors are a good indicator of the amount of productivity required for a given software project when no historical datasets are available [18]. The most interesting finding is that a stronger association between productivity and UCP was found in the datasets that come from industry, such as Industrial. On the other hand, we see a negative correlation between productivity and UCP in datasets containing student or university projects. This may be due to the quality of estimations students make.

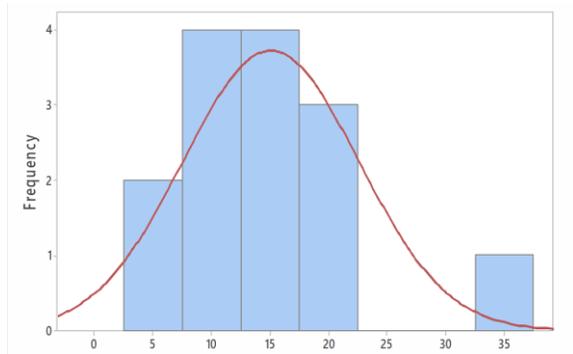
Fig. 4: Productivity Histogram for the Ochodek dataset

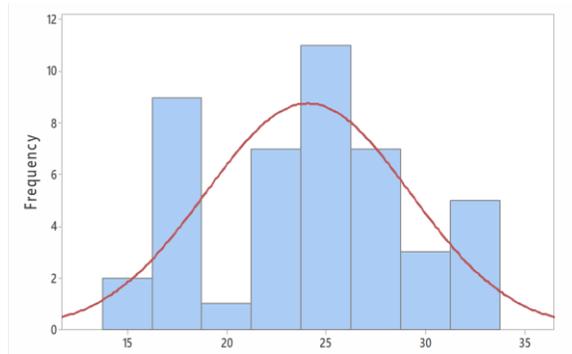
Fig. 5: Productivity Histogram for the Industrial dataset

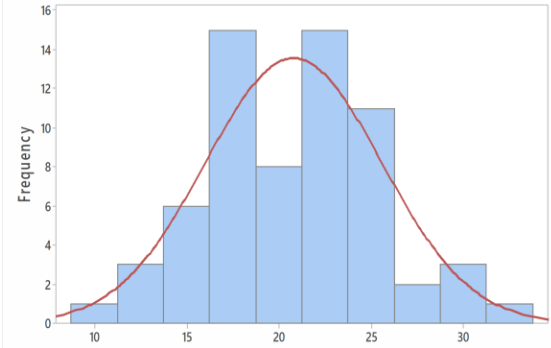

Fig. 6: Productivity Histogram for the Education dataset

Figures 7, 8 and 9 show the relationship between effort and UCP over all datasets. It is clear from Figures 7 and 8 that the relationship is quite linear, with a strong correlation between the two factors in the Ochodek and Industrial datasets. This illustration indicates that UCP is also a good predictor of effort. On the other hand, the relationship between effort and UCP did not show a strong correlation in the Education dataset. Based on these findings, we can see that Equation 6 can work well with the Ochodek and Industrial datasets because the relationship is strong and linear, but not for the Education dataset.

Table 4: Pearson's Rank Correlations between productivity, UCP, effort and EF

|  | Ochodek | | Industrial | | Education | |
| --- | --- | --- | --- | --- | --- | --- |
|  | $r$ | $p$-value | $r$ | $p$-value | $r$ | $p$-value |
| EF | -0.138 | 0.638 | 0.601 | 0.000 | 0.505 | 0.000 |
| UCP | -0.122 | 0.679 | 0.360 | 0.015 | -0.424 | 0.000 |
| Effort | 0.425 | 0.130 | 0.391 | 0.008 | 0.515 | 0.000 |

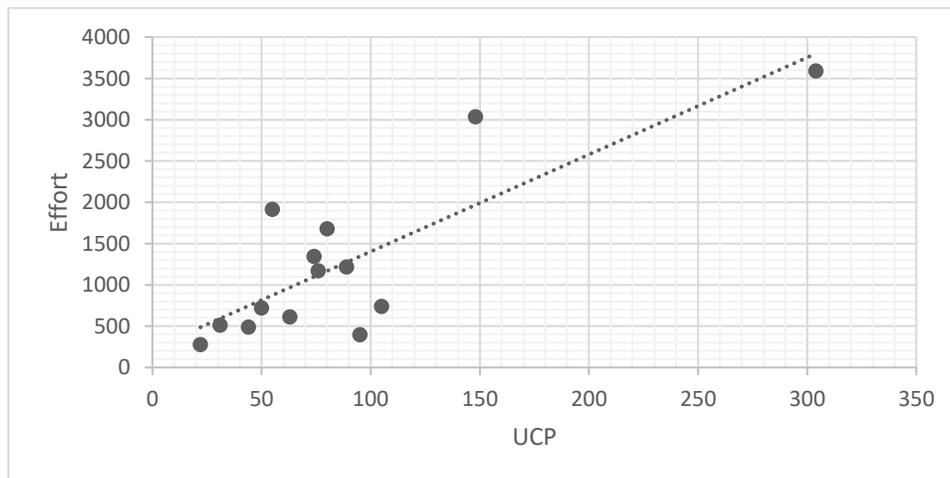

Fig. 7: Relationship between Effort and UCP in the Ochodek dataset (corr=0.82, p-value=0.01)

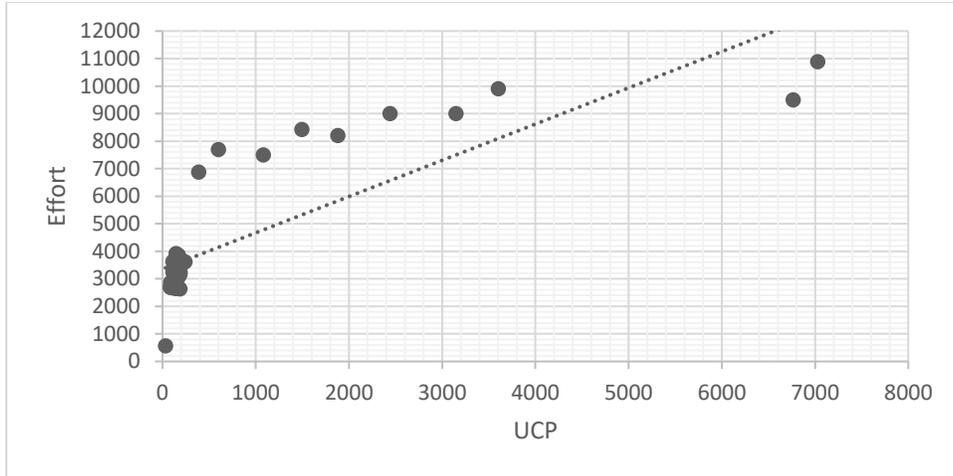

Fig. 8: Relationship between Effort and UCP in the Industrial dataset (corr=0.89, p-value=0.01)

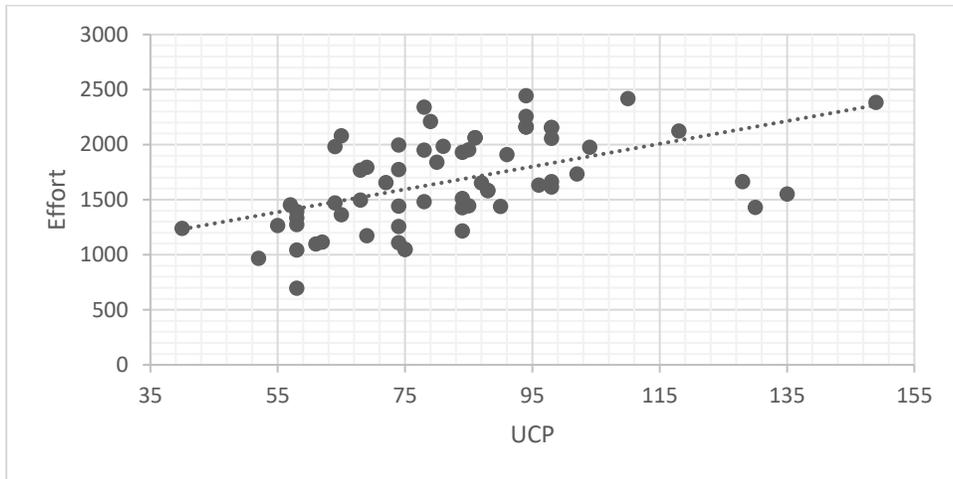

Fig. 9: Relationship between Effort and UCP in the Education dataset (corr=0.52, p-value=0.01)

## 6. Empirical evaluation using productivity to estimate effort

This section presents the empirical results obtained regarding the effectiveness of using productivity to generate effort estimates from UCP. Experiments were performed on three datasets. In this study, we used four different models that use productivity as the factor in estimating effort based on UCP. We first examined whether the models used in the study are predictive or guessing. To achieve this, we used the validation framework proposed by Shepperd and MacDonell based on Standardized Accuracy (SA) and effect size [63]. This framework determines whether the prediction model under investigation produces meaningful predictions (i.e. that is, whether or not the results it produces are significantly better than random guessing). Models with larger SA and effect size are considered as predictive models, while in other models, accurate predictions are likely the result of chance. For each dataset, we report the SA and effect size as shown in Table 5. The results of our SA measurements show that all models produced estimates that proved to be more accurate than random guessing on all datasets, even for the Education dataset. These improvements in estimation are supported by large recorded effect sizes, confirming that these improvements did not occur by chance. For the Ochodek dataset, we can see that the R2M model had the most improvement, while the S&W model had the least improvement. The R2M effect size is greater than one, which means that the difference between the R2M model estimates and random guessing is greater than one standard deviation. For the Industrial dataset, we observe improvements across all models with respect to SA, but the effect size is relatively average. The effect size test produced significantly large effect sizes over all datasets, confirming a significant improvement in guessing (i.e., $\Delta \approx 0.5$). Finally, for the Education dataset, all models predict rather than random guessing, resulting in a very large effect size (greater than three standard deviations). Interestingly, the Karner model, which uses a fixed productivity value for all projects regardless of their complexity and environment, appears to provide stable data and produce meaningful prediction results on most datasets. Similarly, the R2M model tends to produce meaningful predictions with reasonable effect size on all datasets. Based on the results of the size test, we can be sure that all models predict well and achieve more accurate results than can be achieved by random guessing.

Table 5: SA and effect size analysis considering random guessing model as baseline.

| Dataset | | Karner | S&W | Nassif | R2M | Naïve |
|---|---|---|---|---|---|---|
| Ochodek | SA | 0.57 | 0.42 | 0.55 | 0.74 | 0.63 |
| | Δ | 0.93 | 0.54 | 0.82 | 1.36 | 0.93 |
| Industrial | SA | 0.72 | 0.82 | 0.80 | 0.89 | 0.84 |
| | Δ | 0.42 | 0.49 | 0.47 | 0.54 | 0.48 |
| Education | SA | 0.80 | 0.73 | 0.80 | 0.80 | 0.78 |
| | Δ | 3.91 | 2.76 | 3.90 | 3.97 | 3.40 |

Table 6: SA and effect size analysis considering Karner model as baseline.

| Dataset | | S&W | Nassif | R2M | Naïve |
|---|---|---|---|---|---|
| Ochodek | SA | -0.33 | -0.05 | 0.39 | 0.26 |
| | Δ | 0.21 | 0.04 | 0.44 | 0.31 |
| Industrial | SA | 0.36 | 0.28 | 0.62 | 0.19 |
| | Δ | 0.17 | 0.12 | 0.30 | 0.24 |
| Education | SA | -0.36 | 0.01 | 0.03 | -0.05 |
| | Δ | 0.32 | 0.01 | 0.04 | 0.12 |

Similarly, we performed another analysis based on SA and effect size but taking the Karner's model as the baseline model for comparison rather than random guessing. The purpose of this analysis was to record the improvement that could be achieved when using other models over the original Karner model. Table 6 shows the SA and the effect size of the three models used in the study on all datasets. Interestingly, all models show estimation improvements only in the Industrial dataset, and the extent of the improvement can be considered acceptable, but not high. A negative SA means that the model in question did not outperform the Karner model. So, we can determine that all models behave similarly across datasets containing student projects and differently from datasets containing industrial projects. The S&W model was not able to generate better effort estimates than the Karner model when using the Ochodek and Education datasets. The Nassif model improved compared to the Karner model, but only when using the Industrial dataset. The R2M model outperformed the Karner model with the Ochodek and Industrial datasets. No model outperformed Karner using the Education dataset. This raises a concern about the quality of the data contained in these datasets, especially in the Education dataset, which is made up of student projects. One hypothesis is that the students who design the projects in this dataset do not have solid experience translating use case diagrams into UCP as industrial experts do. However, this type of data is still useful to help us understand the shortcomings of this sizing approach and how they can be mitigated. Finally, the most interesting point derived from these is the superiority of the R2M model in dynamically predicting and adjusting productivity from previous projects.

Tables 7 to 9 show accuracy results for all models in all datasets with respect to the three evaluation measures MAE, MBRE and MIBRE. The cells with boldface font represent the best values in the category. The overall results indicate that the R2M model performs better than the other variants for estimating effort based on productivity. Given this comparison, we can conclude that both the R2M and Nassif models perform better than the Karner model, and that the R2M in particular performs better than the S&W model. Another interesting observation is that all prediction models perform well across datasets compiled of data from organizations (industrial datasets). The Karner and S&W models produce the worst performance compared to the R2M. The S&W model produces the worst estimates based on the three datasets and produces a large error deviation, as confirmed by the MAE results. As expected, there is a significant difference in productivity across the different environments. Projects developed in the university are more productive and easier than those developed by organizations.

Table 7: MAE results

| Dataset | Karner | S&W | Nassif | R2M | Naïve |
|---|---|---|---|---|---|
| Ochodek | 639.2 | 850.1 | 669.1 | **388.2** | 481.2 |

| Industrial | 6120.6 | 3893.7 | 4419.9 | **2335.7** | 4943.3 |
| Education | 329.0 | 446.4 | 325.6 | **320.4** | 376.9 |

Table 8: MBRE results

| Dataset | Karner | S&W | Nassif | R2M | Naïve |
|---|---|---|---|---|---|
| Ochodek | 0.79 | 0.88 | 0.67 | **0.49** | 0.60 |
| Industrial | 0.28 | 0.21 | 0.25 | **0.16** | 0.36 |
| Education | 0.24 | 0.30 | 0.24 | **0.23** | 0.29 |

Table 9: MIBRE results

| Dataset | Karner | S&W | Nassif | R2M | Naïve |
|---|---|---|---|---|---|
| Ochodek | 0.34 | 0.37 | 0.31 | **0.24** | 0.29 |
| Industrial | 0.21 | 0.17 | 0.18 | **0.12** | 0.23 |
| Education | 0.18 | 0.20 | 0.18 | **0.17** | 0.19 |

To support this hypothesis, we ran a Mann-Whitney U test for statistical significance on the models' residuals, using an overall significance level of 0.05. The non-parametric Mann-Whitney U test was chosen because of the non-normal distribution of residuals. The results of this comparison are presented in Table 10, which shows that there is statistical significance between Karner's model and other models in most of the datasets. Remarkably, we can see that all models are not statistically different when using the Ochodek dataset except for the R2M model, which shows statistical significance against the Karner and S&W models. In contrast, we can see that all models generate statistically different predictions when using the Education dataset. In fact, we can conclude that the R2M model achieves the best results in terms of the number of wins over other models. This is due to the ability of the R2M model to regress the new estimate to the average productivity of all training projects.

Table 10: The p-values for statistical significance test results between every pair of models

|  | Ochodek | Industrial | Education |
|---|---|---|---|
| Karner vs. S&W | 0.76 | 0.00 | 0.00 |
| Karner vs. Nassif | 0.26 | 0.04 | 0.01 |
| Karner vs. R2M | 0.04 | 0.00 | 0.02 |
| Karner vs. Naive | 0.06 | 0.04 | 0.03 |
| S&W vs. Nassif | 0.12 | 0.19 | 0.00 |
| S&W vs. R2M | 0.03 | 0.02 | 0.00 |
| S&W vs. Naïve | 0.01 | 0.02 | 0.01 |
| Nassif vs. R2M | 0.48 | 0.29 | 0.00 |
| Nassif vs. Naive | 0.01 | 0.01 | 0.03 |
| R2M vs. Naive | 0.03 | 0.01 | 0.04 |

Figures 10 to 13 show the relationship between the actual productivity of a given dataset (*y*-axis) and its estimated productivity (*x*-axis) produced by each of the three selected models—S&W, Nassif, Naïve and R2M—over all projects from three datasets. The productivity of the Karner model is not included at this stage because it has a fixed value (i.e., productivity=20) and therefore has no distribution. The straight line represents the best regression line that passes through all points. However, based on this analysis, the productivity values estimated by the R2M model are less scattered and are closer to the actual productivity, indicating that R2M is more accurate than other models. Figure 10 shows that productivity can be clustered into three groups. As expected from the literature, the relationship between actual productivities and those estimated by Nassif are negatively correlated and do not follow a linear pattern. Nassif et al. [6] confirmed this in their study and showed that non-linear regression is more suitable for effort estimation based on UCP as shown in Figure 11. On the other hand, the scatterplot for the actual productivity and the productivity estimated by the R2M model shows a positive and linear relationship between the two, which means that R2M may be a good predictor for effort estimation based on UCP as shown in Figure 12. Finally, there is no clear correlation between the actual productivity and the estimated productivity for Naïve model as shown in Figure 13.

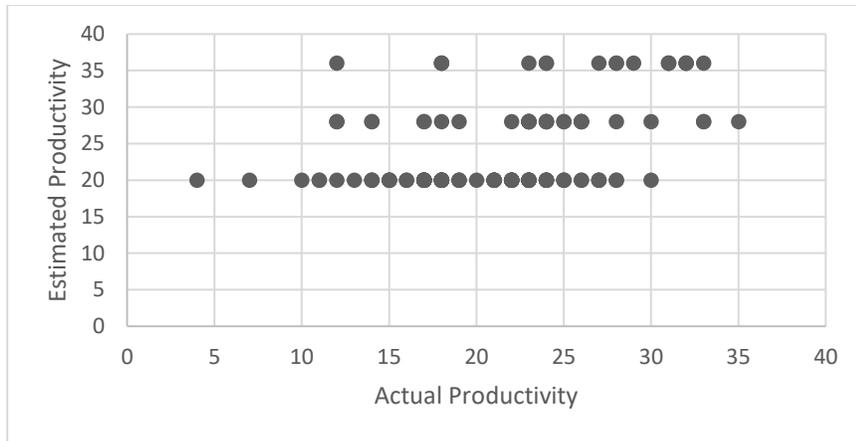

Fig. 10: Scatter plot of the relationship between actual productivity and estimated productivity using the S&W model

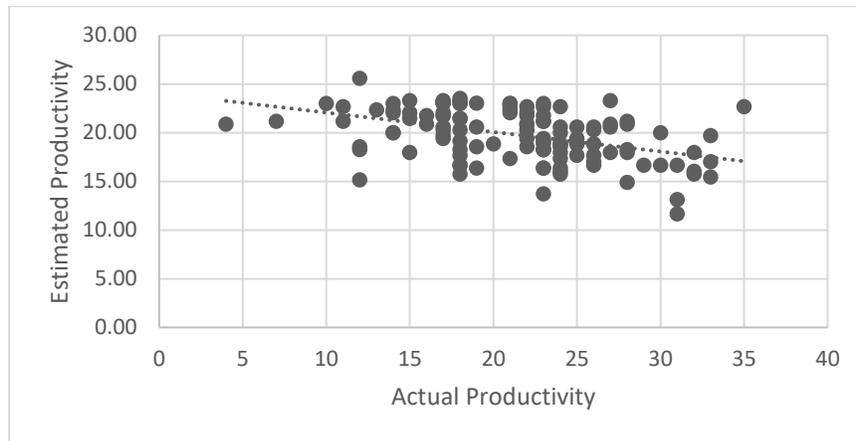

Fig. 11: Scatter plot of the relationship between actual productivity and estimated productivity using the Nassif model

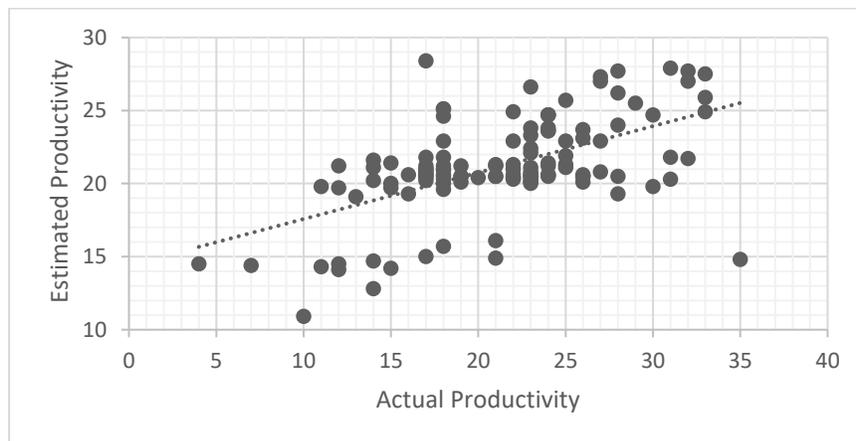

Fig. 12: Scatter plot of the relationship between actual productivity and the estimated productivity using the R2M model

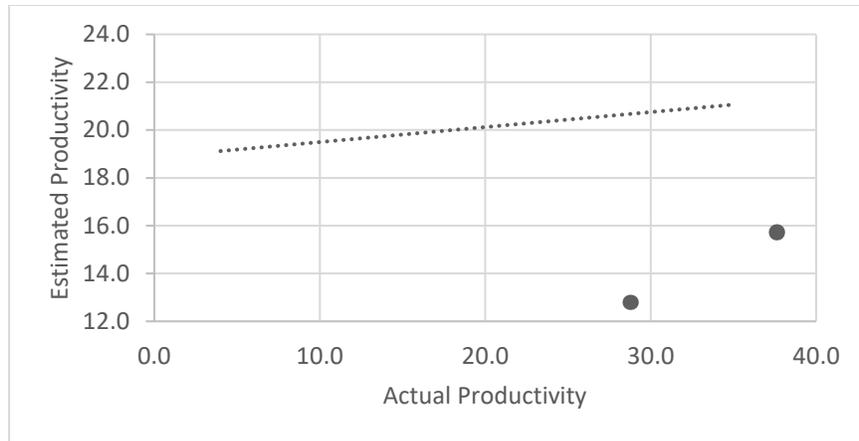

Fig. 13: Scatter plot of the relationship between actual productivity and the estimated productivity using the Naïve model

Figure 14 shows the interval plot at the 95% confidence interval for actual productivity and estimated productivity by S&W, Nassif, Naïve and R2M models across all projects. The interval centers correspond to the average of the estimated productivity of the corresponding model. Models with large interval plot widths indicate a large degree of productivity variability. Based on this analysis, we see that all estimated productivities are statistically significant, which means that each model behaves differently. Moreover, we observe statistical significance between actual and estimated productivity by S&W and Nassif models. In contrast, we could not find any statistical difference between the productivity obtained by the R2M, Naïve models and the actual productivity, which means that the R2M model produces estimates very close to the real productivity of a given project.

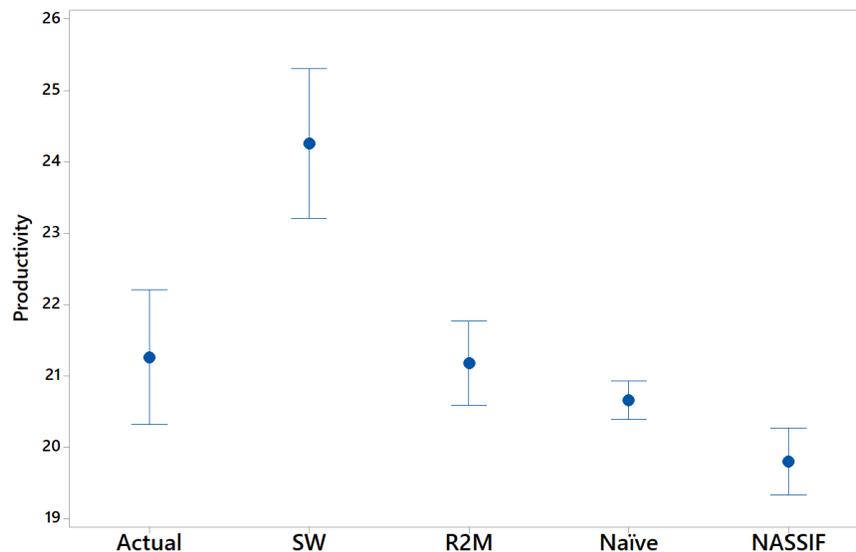

Figure 14: Interval Plot of actual and estimated productivity

## 7. Discussion

This section discusses the findings obtained and provides answers to our research questions. It has already been demonstrated that productivity is main driver in generating early effort estimates based on UCP. The main challenge with this baseline model was estimating productivity both with and without access to the historical dataset. From the literature, we selected four prediction models based on UCP. Two of them require no historical data—the Karner and S&W models—while the other two models require historical data to adjust and tune productivity estimates. We performed several analyzes and empirical experiments on three datasets. Based on the experiments performed, we found that all models included in the study produce reliably and more accurate predictions than random chance, especially when using the Industrial dataset, as shown in Table 5. However, the S&W model performed significantly worse when using a dataset sourced from student projects. This may be due to the uncertain evaluation of environmental factors, which have a significant impact on the S&W model. The model's accuracies in terms of MAE, MBRE and MIBRE are also acceptable, but the R2M model clearly did the best, as confirmed in Tables 7, 8 and 9. Karner and S&W models yielded significantly different predictions when using Industrial and Education datasets, but not when using the Ochodek dataset, which confirms that both models behave differently using homogenous datasets. Here we revisit the research questions and provide answers based on the results obtained from our experiments.

**RQ1**: Can learning productivity values from historical datasets improve the accuracy of UCP effort estimation?
From the results, we observed that the models based on historical data perform better than models based primarily on UCP environmental factors. This is intuitive, because learning from training projects generally enhances prediction models, allowing them to generate meaningful predictions. We also noted that all models perform better when using industrial datasets. The interval plot in Figure 13 indicates that the estimated productivity by the R2M model is very close to the actual productivity. All results were subjected a statistical significance test to ensure that the difference between the models was statistically significant. Based on this, we determined that the R2M model predictions are the most accurate of all the models used in the study. So, we can conclude that the R2M model is best suited for industrial projects, as it outperforms the other models in terms of MAE, MBRE and MIBRE, and is very close to the other models when using a dataset consisting of student projects.

We have seen in Tables 7 through 9 that the models that use different productivity values for each individual project are more accurate than those that use fixed or limited productivity ratios. If we look at Figures 10 to 13, we can see that the relationship between actual and estimated productivities are closer and more linear in the R2M and Nassif models than in the S&W model. Based on these findings, we can confirm that measuring and adjusting productivity based on historical projects is more effective and accurate.

**RQ2**: Is there a strong relationship between productivity and summation of environmental factors?
Previous studies by S&W and Nassif assume that productivity can be predicted by directly summing the effects of Environmental factors (i.e., $EFactor = \sum_{i=1}^{8}(env_i \times w_i)$), but not from EF. We therefore propose to build a linear regression model to predict the productivity from the *EFactor* based on the available data. In this case, the *EFactor* is not included in the UCP size calculation; instead, it will be used to predict productivity. We then multiply the predicted productivity with the UCP to produce an estimation of effort as shown in Equation 6. Before moving forward with further experiments, we ran Dixon's Q outlier test to check if there were any outliers in the EFactor or productivity variables. Accordingly, we found only two outliers in the Ochodeck dataset, which were then excluded from the regression model.

Figures 15 to 17 show the relationship between productivity and the *EFactor* across all datasets. Only the Ochodek dataset provided a positive relationship, while the others provided negative correlations. Negative correlations are usually present in real cases because as the *EFactor* increases, the productivity needed to complete the project decreases. A high *EFactor* usually means that the team members are familiar with the project's development, so high productivity is not required to finish the project. In the case of the Ochodek dataset, the quality of the project data may play an important role in obtaining such an unexpected relationship. It is important to note that when the Ochodek dataset is combined with other datasets, the correlation value decreases due to the fair effect of projects in the Ochodek dataset.

For each dataset we construct a general regression model using productivity and *EFactor,* which may vary marginally on each cross-validation (Leave One Out). General productivity regression models for all datasets are shown in Equations 16 to 18. Since the linear regression model temporally assumes that the predictor and response variables are normally distributed, we must ensure that these variables are transformed to be normally distributed. We performed a normality test and found that both productivity and *EFactor* are normally distributed, so there was no need to convert it to another scale. From the equations, we can see that the adjusted $R^2$ of the Ochodek model is very weak, with only 8% of the productivity values explained by the *EFactor*. Other productivity regression models show a slight improvement, but not a significant one. So, tentatively, we can judge that the productivity prediction from *EFactor* is not feasible and not accurate.

$$\text{Ochodek:} \quad productivity = 6.4 + 0.405 \times EFactor \quad \text{Adj. } R^2 = 8.0\% \quad (16)$$
$$\text{Industrial:} \quad productivity = 4.10 - 0.335 \times EFactor \quad \text{Adj. } R^2 = 34.0\% \quad (17)$$

Education: $productivity = 3.57 - 0.219 \times EFactor$    Adj. $R^2$= 22.2%    (18)

The accuracy results from this approach, after computing efforts from the predicted productivity, are shown in Table 11. Although the new approach leads to some improvement for S&W, Karner and Nassif models on some datasets, it cannot produce better performance than the R2M model with respect to accuracy measures. In general, the new approach tends to produce good estimates with reasonable accuracy over all datasets, so we can safely conclude that the regression model between the productivity and $EFactor$ can produce much better estimates at an early stage.

Table 11: Predictive accuracy of effort estimation based on Equations 16 to 18

| Dataset | MAE | MBRE | MIBRE |
|---|---|---|---|
| Ochodek | 332.6 | 47.5 | 25.5 |
| Industrial | 324.6 | 23.0 | 16.8 |
| Education | 393.0 | 27.9 | 18.7 |

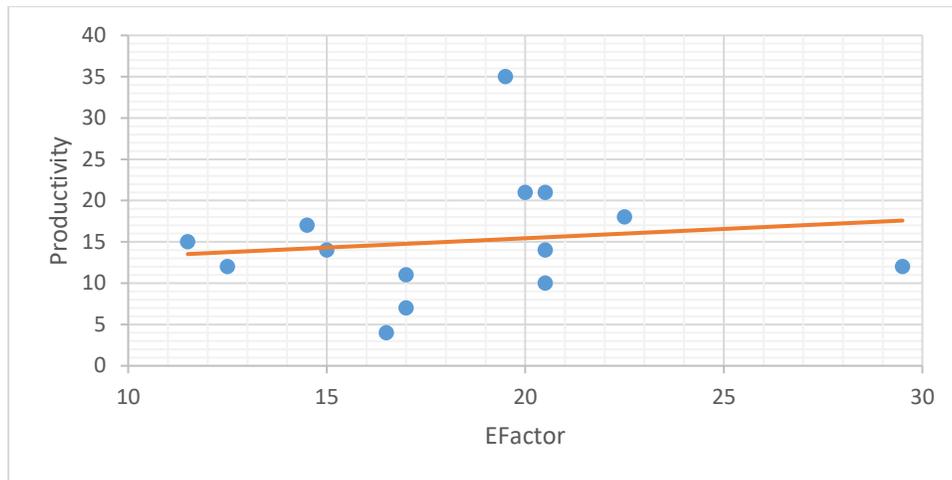

Figure 15. Relatioship between productivity and $EFactor$ over Ochodek dataset (corr=0.284, p-value=0.37)

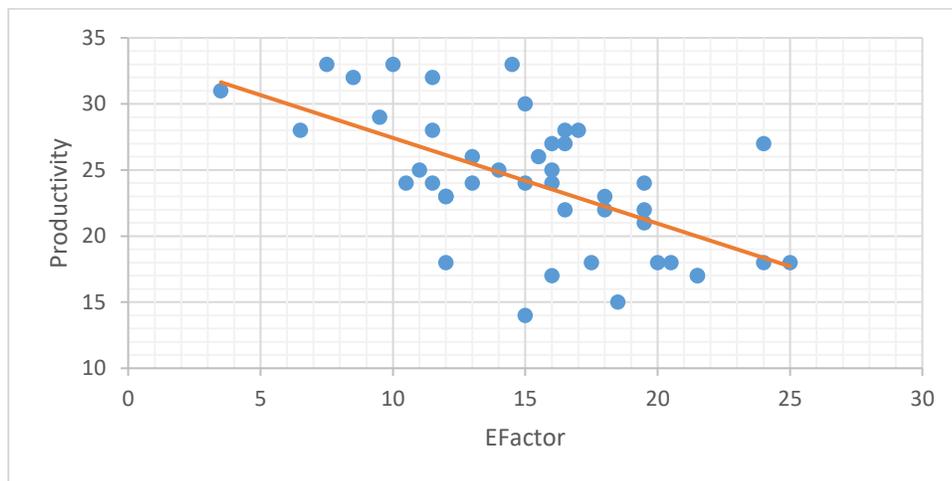

Figure 16. Relatioship between productivity and $EFactor$ over Industrial dataset (corr=-0.61, p-value=0.00)

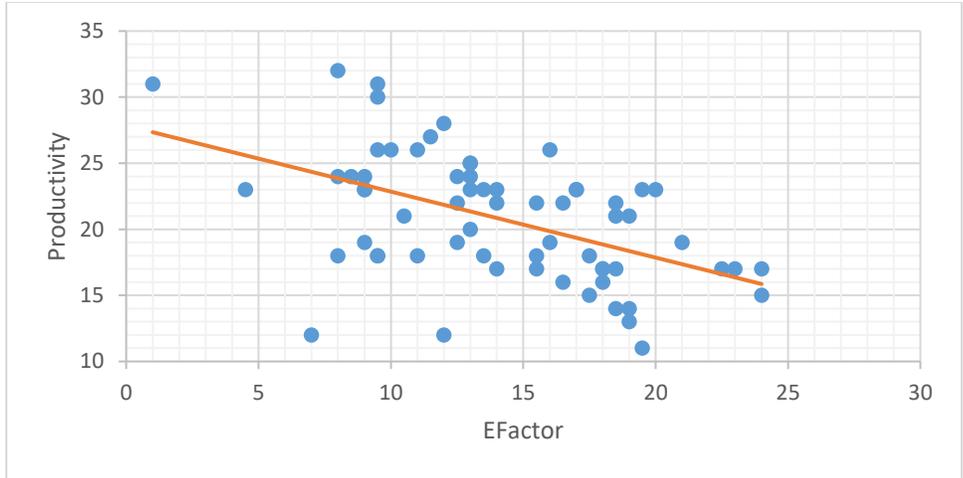

Figure 17. Relatioship between productivity and $EFactor$ over Education dataset (corr=-0.51, p-value=0.00)

**RQ3**: Does using a homogeneous data set improve the accuracy of estimating productivity?

The use of homogenous or heterogeneous datasets has a significant impact on constructing a prediction model. Effort estimation models are easily influenced by the structure of the input data. Therefore, it is important to investigate the stability of predictions in terms of homogenous and heterogeneous data. A homogenous dataset refers to data that shares some characteristics. For example, a heterogeneous software dataset can be divided into a set of homogenous datasets based on origin, programming language, project type, etc. This requires a classification variable to help categorize the larger dataset into smaller ones. In the datasets that we used in this study, we can note that the two Industrial and Education datasets are homogenous in terms of origin; other variables cannot be used to divide them. In contrast, we can see that the Ochodek dataset can be divided into different homogenous datasets based on origin (Industrial or University), programming language (Java) and application type (Web). Therefore, we used the same splitting strategy as Ochodek in their study [12], which produced four homogenous datasets (Industrial, University, Web and Java). Each dataset contains at least 50% of the total data, while other homogenous datasets with small numbers of projects are ignored.

A comparison of the prediction models and the four homogenous datasets, as well as the original Ochodek dataset (all), is shown in Figures 18, 19, and 20. Regarding MAE, we can note that none of the homogenous datasets achieved 100% better accuracy compared to a heterogeneous dataset when all prediction models were used. The results followed the same trend for the other measures of accuracy (MBRE and MIBRE). Specifically, the use of Industrial and Web datasets resulted in a better MAE than the heterogeneous dataset. For MBRE, validations of the three homogeneous datasets (Industrial, Web and Java) resulted in better overall accuracy, but only when using different prediction models. For MIBRE, validations of the three homogenous datasets (University, Web and Java) achieved an improvement in accuracy. Remarkably, we can see that the results of the S&W model are no better than those of the other models when using homogeneous datasets, except for the Industrial dataset. In general, models running on homogeneous datasets produced 58% more accurate results than those running on heterogeneous datasets. Models running on homogenous datasets showed a reasonable improvement over those running on heterogeneous dataset, as described below.

Table 12: Improvement when using homogenous data

| Dataset | MAE | MBRE | MIBRE |
|---|---|---|---|
| Industrial | 75% | 75% | 25% |
| University | 25% | 25% | 75% |
| Web | 75% | 75% | 75% |
| Java | 25% | 75% | 75% |

Finally, the Web dataset was found to be the most accurate dataset. However, the use of homogenous datasets for the Ochodek dataset often does not lead to a significant improvement in accuracy in terms of productivity and effort prediction.

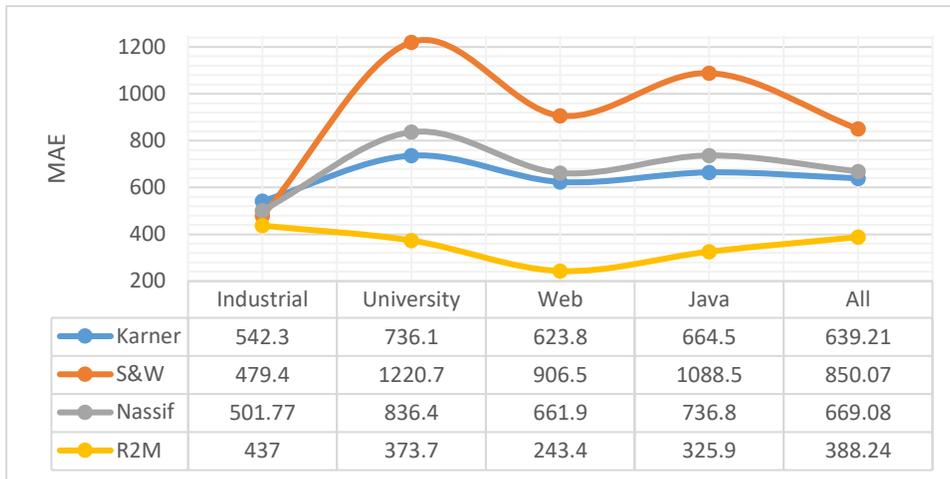

Figure 18: Homogenous and Heterogeneous Data using the Ochodek dataset, measured in MAE

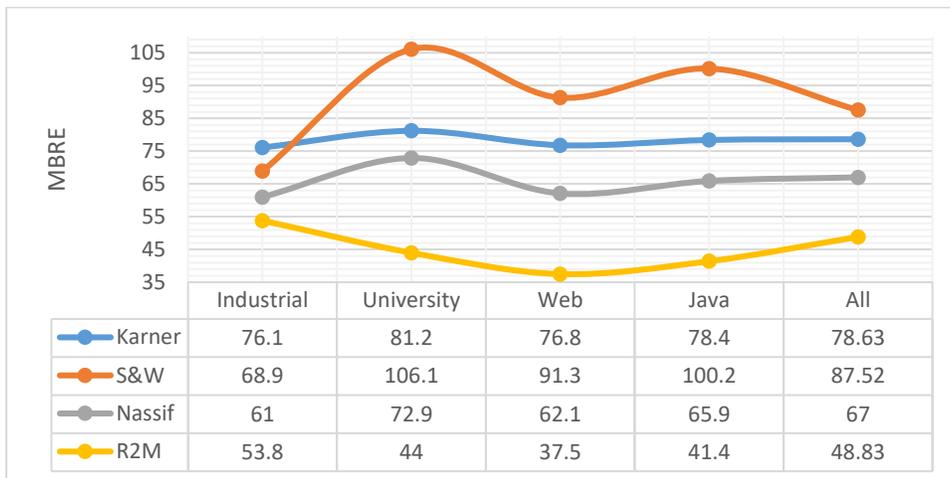

Figure 19: Homogenous and Heterogeneous Data using the Ochodek dataset, measured in MBRE

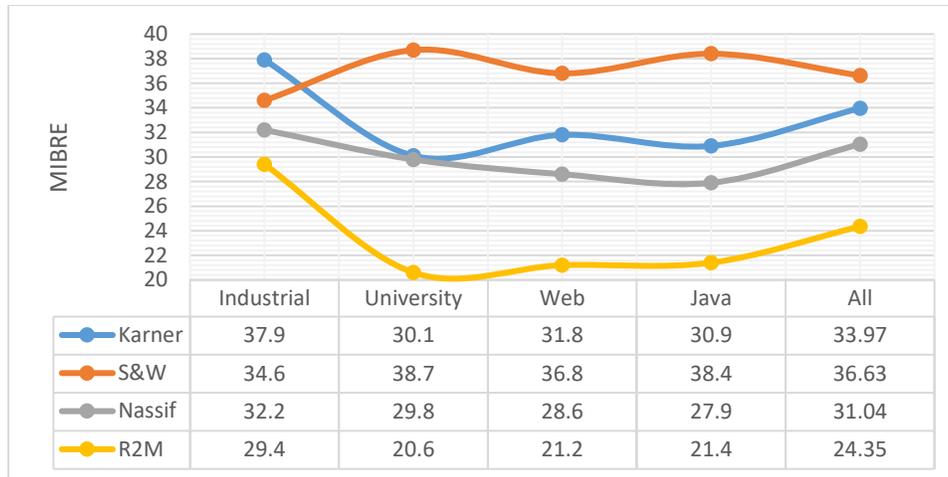

Figure 20: Homogenous and Heterogeneous Data using the Ochodek dataset, measured in MIBRE

## 8. Threats to Validity

This section describes the factors that threaten the validity of our study and present the actions made to alleviate them. We start with internal validity which is the degree to which conclusions can be drawn with regard to the configuration setup. First, the data sampling might have an impact on empirical validation. Although, many different data sampling approaches are available for validating prediction models such k-folds and holdout we favor using leave one out cross validation for two reasons: (1) it produces lower bias estimates and higher variance than n-Fold cross validation [64]. (2) it is a systematic approach for the replicated studies [64]. The choice of productivity models is the second threat. In this study we have chosen the most popular productivity prediction models that have reputation in the field. On the other hand, the construct validity guarantees that we measure what we meant to assess in the first place. Despite the fact that MRE and its related metrics are frequently used to compare prediction models, we did not utilize them in this investigation since they are biased. We employed more robust assessment metrics including SA, effect size, MBRE, and MIBRE instead, which do not cause an unbalanced error distribution. Finally, the external validity is concerned mainly with the dataset's threats. Unlike other areas of effort estimation, UCP effort estimation datasets are still small in number and size. Therefore, the choice of dataset could affect the generalizability of our findings. To overcome this issue, we employed Ochodek dataset that already public and recognized in addition to two quality datasets collected by Nassif et al. [6]. The employed datasets present different characteristics regarding the number of instances and type of projects, which facilitate drawing conclusions. These datasets support our findings and pave the way for generalize our results by other replicated studies.

## 9. Conclusions

This paper examines the significance of learning productivity from historical data in order to estimate effort from UCP. We have conducted a set of empirical experiments using four well-known effort estimation models that use productivity and size variable as predictors. Three datasets were used to perform the experiments. Based on the findings obtained from our empirical experiments, we can summarize the following implications of our research on the software industry:

1. Using different productivity ratios for each individual project seems more effective than using a static ratio for all projects within a given software organization.
2. The different productivity ratios should be learned through a robust model from a historical dataset.
3. Karner and S&W models may work well for educational projects but not for industrial projects.
4. Environmental factors are not good metrics of productivity prediction, but they can be used as a source of data if there are no alternative methods.
5. Environmental factors, as well as technical factors must be carefully evaluated to prevent any possible ambiguity that may cause an inaccurate estimation.
6. The use of homogeneous datasets allows a slight improvement in the productivity and effort estimation for UCP models.

In future work, we plan to study the influence of technical complexity and environmental factors together on the stability of the UCP sizing method. There is great interest in determining the importance of technical and environmental factors for the UCP method. It is necessary to investigate their effect on the accuracy of UCP estimation methods.

## Acknowledgment

Mohammad Azzeh thanks the Princess Sumaya University for Technology for supporting this research.

Yousef Elsheikh is grateful to the Applied Science Private University in Amman, Jordan, for the financial support granted to cover the publication fee of this research article.

Ali Bou Nassif thanks the University of Sharjah for supporting this research